\documentclass[journal,onecolumn,draft,11pt]{IEEEtran}

\usepackage[dvips]{graphicx}
\usepackage{epsfig}
\usepackage[cmex10]{amsmath}
\usepackage{amssymb}
\usepackage{amsthm}
\usepackage{amsfonts}
\usepackage{bm}

\usepackage{cite}
\usepackage[top=0.5in,bottom=0.5in,left=0.625in,right=0.625in]{geometry}
\usepackage[svgnames]{xcolor}
\usepackage[tight,footnotesize]{subfigure}
\usepackage{algpseudocode}

\usepackage{algorithm}

\graphicspath{{figs/}}

\newcommand{\defeq}{\mathrel{\triangleq}}

\newcommand{\floor}[1]{\lfloor{#1}\rfloor}

\newcommand{\iid}{i.\@i.\@d.\ }

\newtheorem{lemma}{Lemma}

\newtheorem{theorem}[lemma]{Theorem}
\newtheorem{corollary}{Corollary}

\theoremstyle{definition}

\newtheorem{egdummy}{Example}

\newtheoremstyle{myremark}
{\topsep}{\topsep}{\normalfont}{\parindent}{\itshape}{:}{ }{}

\theoremstyle{myremark}

\newcounter{Remark}
\newenvironment{Remark}
{
	\refstepcounter{Remark}
	\textbf{Remark \theRemark:}
}

\newcounter{Algorithm}

\newcommand\shortintertext[1]{
	\ifvmode\else\\\@empty\fi
	\noalign{
		\penalty0
		\vbox{\mathstrut}
		\penalty10000
		\vskip-\baselineskip
		\penalty10000
		\vbox to 0pt{
			\normalbaselines
			\ifdim\linewidth=\columnwidth
			\else
			\parshape\@ne
			\@totalleftmargin\linewidth
			\fi
			\vss
			\noindent#1\par}
		\penalty10000
		\vskip-\baselineskip}
	\penalty10000}

\makeatletter
\def\endthebibliography{
	\def\@noitemerr{\@latex@warning{Empty `thebibliography' environment}}
	\endlist
}
\makeatother

\begin{document}
	
\title{Information-Theoretic Secure and Private Voting System}

\author{Seyed Reza Hoseini Najarkolaei, Narges Kazempour, Hasti Rostami, Mohammad Reza Aref\\
	
	Information Systems and Security Lab (ISSL) \\
	Department of Electrical Engineering, Sharif University of Technology, Tehran, Iran
	
}

\maketitle
\begin{abstract} 
 In this paper, we present a private voting system that consists of $N$ authorized voters who may vote to one of the $K$ candidates or vote abstain. Each voter wants to compute the final tally while staying private and robust against malicious voters, who try to gain information about the vote of the other voters beyond the final result, or send incorrect information to affect the final tally.
We design an information-theoretic private voting system based on Shamir secret sharing, which is secure and robust as long as there are up to $\floor{\frac{N-1}{3}}$ malicious voters. 

\end{abstract}

\section{Introduction}
\label{sec:introduction}
The history of voting goes back to ancient Greece, where kings used voting and consensus on a variety of subjects. Over the years, different methods were introduced to satisfy the prerequisites
and constraints in voting. Nowadays, with developments in technology, using electronic voting as an alternative for traditional paper voting has been raised, which is more efficient considering time and resources.

With growing concern about the security and privacy of electronic voting, various protocols and solutions are proposed that satisfy different constraints using diverse tools \cite{DBLP:conf/crypto/Chaum82,DBLP:conf/asiacrypt/FujiokaOO92,ibrahim2003secure,DBLP:journals/cacm/Chaum81,DBLP:conf/ccs/BonehG02,DBLP:conf/trustbus/AdityaLBD04, DBLP:conf/eurocrypt/CramerGS97,chow2008robust,li2013viewable,ayed2017conceptual,ometov2020overview,ccabuk2020survey,binu2016secret}.

One of the tools that
can be used in the voting systems is the blind signature \cite{DBLP:conf/crypto/Chaum82,DBLP:conf/asiacrypt/FujiokaOO92,ibrahim2003secure}. 
The basis of the protocols based on the blind signature is that the authority signs the ballots blindly, and then each voter publishes its ballot through an anonymous channel, and thus, privacy is preserved. Some of the electronic voting protocols use Mix-nets to satisfy privacy \cite{DBLP:journals/cacm/Chaum81,DBLP:conf/ccs/BonehG02,DBLP:conf/trustbus/AdityaLBD04}.
These protocols use shuffle agents to mix the votes, therefore, the authority is unable to find the relationship between the voter and the vote.  Also, there are  electronic voting protocols that use the features of the homomorphic encryption \cite{DBLP:conf/eurocrypt/CramerGS97,chow2008robust,li2013viewable}. With the recent focus on blockchain technology, it has been widely used in electronic voting to provide secure voting
\cite{ayed2017conceptual,ometov2020overview,ccabuk2020survey}.

The aforementioned tools are cryptographic-based secure systems that satisfy different conditions of the voting systems.
Another tool that is used in the voting system to provide information-theoretic privacy is secret sharing.
\cite{binu2016secret}.
Secret sharing is a process of sharing a secret $s$ with other nodes, such that any fewer than or equal to $t$ of colluding nodes cannot gain any information about secret $s$, while any subset of nodes more than some threshold, $t$, can recover it. It was first introduced by Shamir \cite{shamir1979share} which is widely used in the context of coded computing to preserve privacy \cite{ben1988completeness,fullproof,najarkolaei2020coded,nodehi2018entangled,nodehi2019secure,hoseini2020coded}. 
To the best of our knowledge,
the existing information-theoretic voting protocols cannot handle adversarial behavior.

In this paper, we propose an information-theoretic secure formulation for the private voting problem that is robust against adversarial behavior. We consider a private voting problem consisting of $N$ authorized voters that up to $t$ of them are malicious, and $K$ candidates. Each voter may vote to one of the candidates or vote abstain.

The voters are interested in computing the final tally. To do this, voters can interact with each other. The objective is to propose a scheme such that the voters can derive the final tally correctly in the presence of malicious voters, while the privacy is preserved. Note that the malicious voters try to gain information about the vote of the other voters beyond the final result or send incorrect information to affect the final tally.

In the private voting system, three constraints must be satisfied: all valid votes must be counted correctly (correctness), the voters must remain oblivious to the vote of the other voters even if up to $t$ of them collude (privacy), and the system must be robust to malicious behavior of the voters who want to affect the final tally (robustness).

As a solution to the above problem, we propose an information-theoretic private voting system (PVS) utilizing verifiable secret sharing \cite{chor1985verifiable} and multi-party computation \cite{ben1988completeness,hoseini2020coded}. Using verifiable secret sharing ($\mathsf{VSS}$) enables voters to share their votes as a secret such that the privacy of the votes is preserved and other voters can verify the consistency of distributed shares. Also, multi-party computation alongside $\mathsf{VSS}$ enables voters to detect and correct adversarial behavior and compute the final tally, correctly.
The proposed achievable scheme satisfies correctness, privacy, and robustness as long as $N\geq 3t+1$.

The rest of the paper is organized as follows. In Section~\ref{sec:Problem Setting}, we introduce the problem setting. In Section~\ref{sec::PVS}, the main result is represented and some preliminaries are provided in Section~\ref{sec:preliminaries}. We illustrate the motivating example in Section~\ref{section:motivating} and the achievable scheme is proposed in Section~\ref{section:privatesecure}. We conclude the paper in Section~\ref{sec::conclusion}.

\textbf{Notation:}
In this paper matrices and vectors (non-scalar variables) are shown by boldface letters. We show th element-wise multiplication of two vectors $\mathbf{A}$ and $\mathbf{B}$ by $\mathbf{A}*\mathbf{B}$.  $\mathbf{e}_k$ is a vector in $\mathbb{F}^n$ whose components are all zero, except the $k$-th one that is equal to 1. For each $N \in \mathbb{N}$, $[N]$ represents the set $\{1,2,...,N\}$ and $X_{[N]}=\{X_1,X_2,\dots,X_N\}$. Also, for each $\mathbf{V} \in \mathbb{F}^n$, the L-$1$ norm of $\mathbf{V}$ is denoted by Sum($\mathbf{V}$) which is equal to $\displaystyle\sum_{i=1}^{n}|V_i|$. Furthermore,  transpose matrix of $\mathbf{V}$ is shown by $\mathbf{V}^T$ which is yielded by switching its rows with its columns. $\mathbf{1}_{n}$ is a vector in $\mathbb{F}^{n}$ whose components are all one and similarly, $\mathbf{0}_{n\times 1}$ is a null vector in $\mathbb{F}^{n}$.

\section{Problem Setting}
\label{sec:Problem Setting}
The private voting system consists of $N$ authorized nodes $1,2,\dots,N$ as potential voters and $K$ candidates $\mathcal{C}=\{C_1,C_2,...,C_K\}$. Voter $n$ may vote to one of the candidates or vote abstain which is shown by $V_n \in \mathbb{F}^{(K+1)\times 1}$, $\forall n\in[N]$, where $\mathbb{F}$ is a sufficiently large finite field. Voters collaborate with each other and send a function of their vote to the other voters.
The objective is, for each voter to be able to compute the final result of voting $\mathbf{R}=[R_1,R_2,\dots,R_{K+1}]^T$, where
$R_k$ is the tally of casted votes corresponding to candidate $C_k$, $\forall k \in [K]$, and  $R_{K+1}$ shows the number of abstain votes. Also, assume that up to $t$ of the voters are malicious. The malicious voters may send incorrect data to the other voters to affect the final result of voting. Besides, the malicious voters may violate privacy, i.e., they want to get information about the votes of the other voters. To achieve their goals, the malicious voters can collude, share their data with each other, or deviate from the protocol. Note that the voters do not know in advance which of them are malicious. Therefore, one of the challenges for the voters is to compute the final result correctly in the presence of the malicious voters, while keeping their votes private.

In the PVS, each pair of voters are connected to each other with a point-to-point private link. Also, there is an authenticated broadcast channel among all voters such that the identity of the broadcaster is known. All of the links and channel are error-free and secure.

The proposed scheme consists of 3 steps:

\begin{enumerate}
	\item \textbf{Sharing}: In this step, each voter $n \in [N]$ shares its vote, i.e., it sends a function of $\mathbf{V}_n$ to all other voters. Let $\mathcal{S}_{n,n'}\defeq \mathbf{F}_{n,n'}(\mathbf{V}_n)$ be the set of all messages that voter $n'$ received from voter $n$ in this step, where $\mathbf{F}_{n,n'}:\mathbb{F}^{(K+1)\times 1} \rightarrow \mathbb{F}^{p\times q}$, for some $p,q \in \mathbb{N}$ and $n' \in [N]$. For simplicity, let us define $\mathcal{S}_n \defeq \displaystyle\cup_{n'=1}^{N} \mathcal{S}_{n',n} $, be the set of all messages that voter $n$ received in this step. 

	\item \textbf{Verification}: In this step, voters process their input messages from the previous step and communicate with each other to be able to verify the validity of each vote. A vote is valid if it is compatible with the voting system being used, e.g., the vote does not contain additional and surplus entries by the voter or more choice than permitted (overvoting).  In this step, any adversarial behavior can be detected, corrected, or dropped. Let $\mathcal{M}_{n,n'}$ be the set of all messages that voter $n'$ received from voter $n$ in this step, $\forall n,n' \in [N]$ and $\mathcal{M}_n \defeq \displaystyle\cup_{n'=1}^{N} \mathcal{M}_{n',n} $ is the set of all messages that voter $n$ received in this step.
	
	\item \textbf{Counting}: After verification step, each voter $n \in [N]$, broadcasts a message $\mathcal{B}_n$ to all the other voters. By using $\mathcal{B}_1,\mathcal{B}_2,\dots,\mathcal{B}_N$, each voters must be able to derive the final result, correctly.

\end{enumerate}

To achieve the goal of the proposed voting system, the PVS must satisfy three constraints, correctness, privacy, and robustness as follows.

\begin{itemize}
	\space \item	\textbf{Correctness}: 
		All valid votes must be counted correctly, i.e., after the execution of the proposed algorithm, each voter must have sufficient information to be able to derive the final result $\mathbf{R}$, where it must be effectively the real tally of the casted vote. More precisely,
		\begin{align}
		\label{correctness}
		H(\mathbf{R}|\mathcal{S}_{n}, \mathcal{M}_{n},\mathcal{B}_{[N]})=0, \forall n \in [N].
		\end{align}

		Note that the correctness condition must be satisfied in the presence of at most $t$ malicious voters. 
\end{itemize}

\begin{itemize}
	\space \item	\textbf{Privacy}:  If any arbitrary subset $\mathcal{X}$ of at most $t$ voters coalitate, cannot gain any information about the vote of the other voters beyond the final result $\mathbf{R}$. It means that, for each $n \in [N]\backslash \mathcal{X}$, then
	
	\begin{align}
	\label{privacy}
	 H(\mathbf{V}_n|\mathbf{R},\mathcal{S}_{\mathcal{X}}, \mathcal{M}_{\mathcal{X}},\mathcal{B}_{[N]})
	 =
	 H(\mathbf{V}_n| \mathbf{R}).
	\end{align}
	
\end{itemize}

\begin{itemize}
	\space \item    \textbf{Robustness}:  Each voter must be able to vote exactly once, and no voter can vote more than once. A voter's vote cannot be changed, duplicated, or removed by malicious voters. Any adversarial behavior of at most $t$ of the voters, can be tolerated. No adversarial treatment can disrupt the voting and any cheating behavior will be detected or corrected.
\end{itemize}

\section{Main Result} \label{sec::PVS}
The objective of PVS is to derive the final tally correctly while staying private and robust.
In this paper, we propose a new private voting scheme explained in Section \ref{section:privatesecure} which is robust against adversarial behavior. The main result is stated in the following theorem.
\begin{theorem}\label{thm::main_result}
Given $K$ candidates and $N$ voters, such that up to $t$ of them are malicious, there exists a private voting scheme that satisfies correctness, privacy, and robustness conditions as defined in Section \ref{sec:Problem Setting}, as long as $N\geq 3t+1$.
\end{theorem}

\begin{Remark}
The achievable scheme is provided in Section~\ref{section:privatesecure}. It is based on $\mathsf{VSS}$ and multi-party computation. $\mathsf{VSS}$ enables voters to share their votes as a secret such that the privacy of the votes is preserved and other voters can verify the consistency of distributed shares. Also, multi-party computation enables voters to detect and correct adversarial behavior and compute the final result correctly.
\end{Remark}

\begin{Remark}
To the best of our knowledge, the proposed achievable scheme is the only information-theoretic private voting system that is robust against adversarial behavior.
\end{Remark}

\begin{Remark}
The minimum number of voters needed depends linearly on the number of malicious voters with a coefficient $3$. The upper bound $3t+1$ is a common phenomenon in distributed computation with malicious nodes. Also, one can see that the number of candidates can not affect $N$ .
\end{Remark}

\begin{Remark}
		In the proposed framework, the voters perform all the computing, and each of them can compute the final result, i.e., the voting is performed completely inside the group of voters. The setting of the problem can be changed to the master-slave framework. In this formation, the voters only send their shares and required data to the workers. Then, the workers perform the computing and send the tally to an authority. The master-slave framework of the voting system can be handled with a slight difference in our proposed achievable scheme.
\end{Remark}

\section{Preliminaries} \label{sec:preliminaries}
Before describing the achievable scheme, we need some preliminaries.
\subsection{Polynomial Interpolation and Reed-Solomon Codes}
\label{subsubsec:Lagrangetheorem}
Constructing a polynomial that passes through a desired set $\mathcal{S}$ of points is called polynomial interpolation. Lagrange theorem which is stated in Theorem \ref{theorem:Lagrange}, is used to find the minimum-degree polynomial that goes through points in $\mathcal{S}$.

\begin{theorem}[Lagrange theorem]
	\label{theorem:Lagrange}
Assume that $x_1,x_2,\dots,x_{t+1}$ are distinct  elements of  $\mathbb{F}$ and $y_1,y_2,\dots,y_{t+1}$ are elements of  $\mathbb{F}$ (not necessarily distinct). There exists a unique polynomials $p(x)$ of degree at most $t$, such that $p(x_i)=y_i$, $\forall i \in [t+1]$.
\end{theorem}

\begin{IEEEproof}
see \cite{werner1984polynomial}.
\end{IEEEproof}

\begin{corollary}
	One can see that, by using the Lagrange theorem, any polynomial of degree $t$ can be uniquely represented by $t+1$ points that lie on it.
\end{corollary}

\begin{Remark}
Suppose that $x_1,x_2,\dots,x_{N}$ are distinct  elements of  $\mathbb{F}$ and $y_1,y_2,\dots,y_{N}$ are not necessarily distinct elements of  $\mathbb{F}$. Also assume that $c$ elements of the set $\mathcal{P}=\{(x_1,y_1),(x_2,y_2),\dots,(x_N,y_N)\}$ are on a polynomial $p(x)$ of degree~$t$, where $N>t$. Hence, Reed-Solomon decoding procedure guarantees that $p(x)$ can be reconstructed by using the points of the set $\mathcal{P}$, if $N-c$ (number of errors) is at most $\floor{\frac{N-t}{2}}$ \cite{wicker1999reed}.
\end{Remark}

\begin{corollary}
\label{remark:reedsolomon}
Consider that $x_1,x_2,\dots,x_{N}$ are distinct  elements of  $\mathbb{F}$ and $y_1,y_2,\dots,y_{N}$ are not necessarily distinct elements of  $\mathbb{F}$. Also, assume that at most $t$ elements of the set $\mathcal{P}=\{(x_1,y_1),(x_2,y_2),\dots,(x_N,y_N)\}$ are not located on a polynomial of degree $t$ called $p(x)$, where $N>t$.
Reed-Solomon decoding procedure guarantees that if $N\geq3t+1$, then $p(x)$ can be reconstructed uniquely by using elements of the set $\mathcal{P}$.
\end{corollary}

\subsection{Verifiable Secret Sharing}
\label{subsubsec:VSS}
Assume that a node in a system, called as \emph{the dealer} wants to share the secret $s$ with other nodes, such that any fewer than or equal to $t$ of colluding nodes cannot gain any information about secret $s$, while any subset of nodes more than some threshold, $t$, can recover it. 
Secret sharing was first introduced by Shamir \cite{shamir1979share1} and Blakley \cite{blakley1979safeguarding}, independently, in 1979.
It is a basic tool in cryptography and has been used in many applications such as e-voting schemes, crypto-currencies, and access control systems. 
Shamir secret sharing is a method by which a secret $s$ can be shared among $N$ participants such that each of the participants has a share of the secret $s$ and a certain number of participants shown by $t$ is required to be able to recover the secret. In the Shamir secret sharing scheme, the dealer who has the secret $s$, constructs a polynomial $f(x)=s+c_1x+c_2x^2+\dots+c_tx^t$ of degree $t$ such that the constant term of $f(x)$ is equal to the secret and the other coefficients are chosen uniformly and randomly from the field $\mathbb{F}$. Assume that each participants $n$ is assigned a distinct and nonzero $\alpha_{n} \in \mathbb{F}$. Then the dealer sends $f(\alpha_n)$ to participant $n$, $\forall n \in [N]$, where $\alpha_{1},\alpha_{2},\dots,\alpha_{N}$ are chosen uniformly and randomly from the field. One can see that any arbitrary subset $\mathcal{X}$ of at least $t+1$ participants can find the secret $s$ in collaboration with each other, but if the size of $\mathcal{X}$ was at most $t$, they can not gain any information about the secret.  It can be shown that this scheme is information-theoretically secure. In this scheme, we assume that the dealer is trusted and always sends \emph{consistent shares} to the other nodes, i.e., it chooses points on a polynomial of degree $t$.

In many cases, the dealer is malicious and may send non-consistent shares to the other nodes. In this case, we need a mechanism that is able to verify the consistency of the shares. Chor et al. \cite{chor1985verifiable} introduce verifiable secret sharing ($\mathsf{VSS}$), which enables nodes to confirm whether their shares are consistent or not. 
The work of \cite{chor1985verifiable} has been followed by many other results, which can be categorized into two major approaches.

\begin{enumerate}
	\item Computational $\mathsf{VSS}$ schemes:
	In those schemes, we assume that adversaries have bounded computing power that limits their ability to solve some mathematical problems with extensive complexity, such as finding prime divisors of a large composite number. Some examples of computational $\mathsf{VSS}$ can be found in \cite{feldman1987practical,pedersen1991non}.
	\item Information theoretically secure $\mathsf{VSS}$ schemes:
	In this case, we do not limit the adversaries in terms of computational power or storage size. Those kinds of schemes are information-theoretically secure, i.e., the security holds, even if the adversary has unbounded computing power, such as \cite{benaloh1986secret,stinson1999unconditionally,patra2009efficient}.
\end{enumerate}

In the context of Shamir secret sharing, verifiable secret sharing  has the following properties:
\begin{itemize}
	\item If the dealer is malicious, and the shares that it sends to the other nodes are not consistent, i.e., are not some points on a polynomial of specified degree, then the honest nodes in collaboration with each other will realize that and reject the shares.
	\item If the dealer is honest, then the malicious workers cannot deceive the honest nodes and convince them that the dealer is malicious; thus, each honest node accepts its share.
\end{itemize}
In its original form \cite{ben1988completeness}, to share a secret $s$ from a field $\mathbb{F}$, the dealer chooses a bivariate polynomial $S(x,y)$, uniformly at random from the set of all bivariate polynomials of degree $t$, with respect to each of the variables $x$ and $y$, with coefficients from $\mathbb{F}$, subject to $S(0,0)=s$. Then, the dealer sends $f_n(x) \defeq S(x,\alpha_n)$ and $g_n(y) \defeq S(\alpha_n,y)$ to the worker $n$,  $\forall n \in [N]$ and some distinct $\alpha_n \in \mathbb{F}$. One can see that, $\forall n,n' \in [N]$, $f_n(\alpha_{n'})=g_{n'}(\alpha_{n})$. Therefore, the redundancy in this scheme allows the honest workers to verify the consistency of shares through communication with other workers.
and it used as follows. For each pair of $n,n' \in [N]$, node $n$ sends $f_n(\alpha_{n'})$ and $g_n(\alpha_{n'})$ to node $n'$. Then node $n'$ is able to verify that whether their univariate polynomials are pairwise consistent, i.e.,  $f_n(\alpha_{n'})=g_{n'}(\alpha_n)$ and $f_{n'}(\alpha_{n})=g_n(\alpha_{n'})$ or not. If it is not, node $n'$ broadcasts a \textsf{complaint} message including $(n',n,f_{n'}(\alpha_{n}),g_{n'}(\alpha_{n}))$. If these values are correct, then the dealer do nothing. else, it broadcast both of $f_{n'}(x)$ and $g_{n'}(x)$
For detailed description of $\mathsf{VSS}$, refer to \cite{fullproof}. 

\section{Motivating Example}
\label{section:motivating}

For ease of understanding, first, we demonstrate the main idea of PVS through a simple example. Consider a private voting system with $N$ voters such that each voter can vote "$\mathsf{Yes}$" or "$\mathsf{No}$". The objective is to derive the total number of "$\mathsf{Yes}$" votes. In this system, "$\mathsf{No}$" vote and "$\mathsf{Yes}$" vote are shown by $0$ and $1$, respectively. The steps of the proposed algorithm are as follows.
\subsection{Sharing} \label{subsection:sharing}
In this step, each voter, shares both $0$ and $1$ using verifiable secret sharing algorithm~\cite{chor1985verifiable}. In order to do that voter $n$ constructs polynomials $F^{(n)}(x)=0+R^{(n)}_{1}x+R^{(n)}_{2}x^2+\dots+R^{(n)}_{t}x^t$ and $G^{(n)}(x)=1+Z^{(n)}_{1}x+Z^{(n)}_{2}x^2+\dots+Z^{(n)}_{t}x^t$, then sends $F^{(n)}(\alpha_{n'})$ and $G^{(n)}(\alpha_{n'})$ to voter $n'$, $\forall n,n' \in [N]$, where $R^{(n)}_{k}$ and $Z^{(n)}_{k}$ are chosen uniformly and independently at random from the field $\mathbb{F}$, $\forall k \in [t]$. Also, distinct $\alpha_{1},\alpha_{2},\dots,\alpha_{N}$ are chosen uniformly and independently at random from the field $\mathbb{F}$ and they are known by all the voters.

Using $\mathsf{VSS}$ \cite{chor1985verifiable} ensures the voters that if $N \geq 3t+1$, shared values by the voter $n$ are consistent, i.e., they are indeed on a polynomial of degree $t$, otherwise, honest (not malicious) voters can identify malicious voters who have adversarial behavior and omit them from the remaining part of our algorithm. It must be mentioned that if voter $n$ is honest, the other voters cannot gain any information about the secret of $F^{(n)}(x)$ and $G^{(n)}(x)$,
i.e., the values $F^{(n)}(0)$ and $G^{(n)}(0)$.
	
\subsection{Verification}
\label{subsection:verification}
In this step, each voter needs to be assured that $\forall n \in [N]$, exactly one of the $F^{(n)}(0)$ and $G^{(n)}(0)$ is equal to $1$ and the other one is equal to $0$. To do this, we perform 2-phase verification. In the first phase, verification of summation, voters verify  whether $F^{(n)}(0) + G^{(n)}(0)$ is equal to $1$ or not, and in the second phase, verification of product, they verify  whether $F^{(n)}(0)G^{(n)}(0)$ is equal to $0$ or not.
If both of the aforementioned conditions are satisfied, then we can conclude that $\{F^{(n)}(0),G^{(n)}(0)\}=\{0,1\}$.

\subsubsection{\textbf{Verification of summation}} Let us define $S^{(n)}(x) \defeq F^{(n)}(x)+G^{(n)}(x)$. In this phase, $\forall n,n' \in [N]$, each voter~$n'$ broadcasts $S^{(n)}(\alpha_{n'})=F^{(n)}(\alpha_{n'})+G^{(n)}(\alpha_{n'})$. If all of the voters were honest, after this phase each voter has access to $\{S^{(n)}(\alpha_{1}),S^{(n)}(\alpha_{2}),\dots,S^{(n)}(\alpha_{N})\}$. But in real, some of the voters are malicious and do adversarial behavior. One can see that $\deg(S^{(n)}(x))=t$, thus, due to the Remark~\ref{remark:reedsolomon},
voters can correct up to $\frac{N-t}{2}$ errors. Since the number of malicious voters who are omitted or not is at most $t$, we need to have $\frac{N-t}{2}\geq t$, or equivalently, $N\geq 3t+1$.
If $N\geq 3t+1$, each voter can recover the correct set of $\{S^{(n)}(\alpha_{1}),S^{(n)}(\alpha_{2}),\dots,S^{(n)}(\alpha_{N})\}$. Thus, each voter can calculate $S^{(n)}(x)$, then, derive $S^{(n)}(0) = F^{(n)}(0)+G^{(n)}(0)$, and verify whether $F^{(n)}(0) + G^{(n)}(0)$ is equal to $1$ or not, $\forall n \in [N]$.
	
\subsubsection{\textbf{Verification of Product}} As it is aforementioned, in this phase, each voter needs to verify whether $F^{(n)}(0)G^{(n)}(0)$ is equal to $0$ or not, $\forall n \in [N]$. This scheme is known as \emph{sharing a product of shares} which is explained in \cite{fullproof}. To be self-contained, the following is a brief overview of the scheme.	First, we express a theorem from Subsection 6.6 of \cite{fullproof}:
	
	\begin{theorem}\cite[Subsection 6.6]{fullproof}
		\label{theorem:Oiha}
		For each pair of arbitrary polynomials $A(x)$ and $B(x)$ of degree $t$, there exist $t$ polynomials $O_1(x),O_2(x),\dots,O_t(x)$ of degree $t$ such that the degree of $A(x)B(x)-\displaystyle\sum_{i=1}^{t}x^i O_i(x)$ is equal or less than $t$. 
	\end{theorem}

According to Theorem \ref{theorem:Oiha}, each voter $n$, can find polynomials $O^{(n)}_1(x),O^{(n)}_2(x),\dots,O^{(n)}_t(x)$, such that $\deg( F^{(n)}(x)G^{(n)}(x)-\displaystyle\sum_{i=1}^{t}x^i O^{(n)}_i(x)) \leq t$. Let us define

\begin{align}
\label{eq:mult}
C^{(n)}(x) \defeq F^{(n)}(x)G^{(n)}(x)-\displaystyle\sum_{i=1}^{t}x^i O^{(n)}_i(x).
\end{align}

One can see that $C^{(n)}(0)=F^{(n)}(0)G^{(n)}(0)$. This is due to the fact that each $O^{(n)}_i(x)$ is multiplied by $x^i$, where $i \geq 1$. Thus, the constant term of $F^{(n)}(x)G^{(n)}(x)$ cannot be affected by $O^{(n)}_i(x)$, $\forall i \in [t]$.
Constructing $C^{(n)}(x)$ enables other voters to compute the value of $F^{(n)}(0)G^{(n)}(0)$ without violating the privacy, i.e., malicious voters cannot get any additional information about the polynomials $F^{(n)}(x)$ and $G^{(n)}(x)$.

After constructing $O^{(n)}_1(x),O^{(n)}_2(x),\dots,O^{(n)}_t(x)$, voter $n$ shares $O^{(n)}_i(x)$ with all other voters by using $\mathsf{VSS}$ algorithm, i.e., it sends $O^{(n)}_{i}(\alpha_{n'})$ to voter $n'$, $\forall n,n' \in [N]$, and  $\forall i \in [t]$. In addition, voter $n$ shares $C^{(n)}(x)$, i.e., it sends $C^{(n)}(\alpha_{n'})$ to voter $n'$, $\forall n,n' \in [N]$. Until now, voter $n'$ has the values of $C^{(n)}(x),F^{(n)}(x),G^{(n)}(x),O^{(n)}_i(x)$ at point $\alpha_{n'}$, $\forall n,n' \in [N]$ and $\forall i \in [t]$. Hence, voter $n'$ can directly verify whether \eqref{eq:mult} is held at $\alpha_{n'}$ or not. If \eqref{eq:mult} is not held at $\alpha_{n'}$, voter $n'$ broadcasts a $\mathsf{Complaint}$ messages. As explained in detail in \cite[Subsection 6.5]{fullproof}, the other voters can compute the values $C^{(n)}(\alpha_{n'}),F^{(n)}(\alpha_{n'}),G^{(n)}(\alpha_{n'}),O^{(n)}_1(\alpha_{n'}),$ $O^{(n)}_2(\alpha_{n'}),\dots,O^{(n)}_t(\alpha_{n'})$ in collaboration with each other, to identify the malicious voter among voter $n$ and voter $n'$ and omit the malicious one from the remaining part of the algorithm.
To be more precise, one can see \cite{fullproof}.

Then, to verify $F^{(n)}(0)G^{(n)}(0)=0$, each voter $n'$ broadcasts the value of $C^{(n)}(\alpha_{n'})$. Hence, each voter has access to the value of $C^{(n)}(x)$ at more than $3t$ points. Thus, due to the Remark~\ref{remark:reedsolomon}, each voter can compute $C^{(n)}(x)$ and verify if the value of $C^{(n)}(0)=F^{(n)}(0)G^{(n)}(0)$ is equal to 0 or not.

\subsection{Counting} \label{subsection:counting}
Assume that $\mathcal{I}$ is the set of all malicious voters that are identified by the other voters. So far, each voter $n'$ has $F^{(n)}(\alpha_{n'})$ and $G^{(n)}(\alpha_{n'})$ , $\forall n,n' \in [N]\backslash \mathcal{I}$. Also it is sure that  exactly one of the $F^{(n)}(0)$  and $G^{(n)}(0)$ is equal to 1 and the other one is equal to 0.  Then, voter $n$ broadcasts which polynomial between $F^{(n)}(x),G^{(n)}(x)$ is its vote. For simplicity,  voter~$n$ broadcasted  polynomial is denoted by $V^{(n)}(x)$ and let us define $V(x) \defeq \displaystyle\sum_{n \in [N]\backslash \mathcal{I}}V^{(n)}(x)$. Voter $n'$ computes $V(\alpha_{n'})= \displaystyle\sum_{n \in [N]\backslash \mathcal{I}} V^{(n)}(\alpha_{n'})$ and broadcasts the result. Ideally, after this step, each voter has access to $\{V(\alpha_{1}),V(\alpha_{2}),\dots,V(\alpha_{N})\}$. But in reality, some of the voters are malicious and do adversarial behavior. One can see that $\deg(V(x))=t$, thus, by using Reed-Solomon decoding procedure, voters can correct up to $\frac{N-t}{2}$ errors. Since the number of malicious voters who are omitted or not is at most $t$, we need to have $\frac{N-t}{2}\geq t$, or equivalently, $N\geq 3t+1$.
If $N\geq 3t+1$, each voter can recover the correct set of $\{V(\alpha_{a_1}),V(\alpha_{a_2}),\dots,V(\alpha_{a_{|[N]\backslash\mathcal{I}|}})\}$, $\forall \alpha_i \in [N] \backslash \mathcal{I}$, and calculate $V(x)$, and finally derive $V(0) = \displaystyle\sum_{n \in [N]\backslash \mathcal{I}}V^{(n)}(0)=\displaystyle\sum_{n \in [N]\backslash \mathcal{I}}V_n$, which is the total number of $1$ casted in our voting without counting the votes of identified malicious voters in set $\mathcal{I}$ .\\
As described above as long as $N\geq 3t+1$, the correctness and robustness properties are satisfied. The privacy is assured using verifiable secret sharing, the detailed proof is provided in Appendix \ref{sec:privacy}.

\section{General scheme}
\label{section:privatesecure}

Consider a voting system consisting of $N$ authorized voters $1,2,\dots,N$ that at most $t$ of them are malicious, and there are $K$ candidates $\mathcal{C}=\{C_1,C_2,...,C_K\}$. Voter $n$ may vote to one of the candidates or abstain which is shown by $V_n \in \{0,1\}^{(K+1)\times 1}$, $\forall n\in[N]$.
The voters aim to compute the final result of voting $\mathbf{R}=[R_1,R_2,\dots,R_{K+1}]^T$, where
$R_k$ is the tally of casted votes corresponding to candidate $C_k$, $\forall k \in [K]$, and  $R_{k+1}$ shows the number of abstain votes.

In this section, we will follow the same protocol as Section \ref{section:motivating} with some modifications to handle more candidates. The steps of the proposed algorithm are as follows.
\subsection{Sharing}
\label{subsection:PVSsharing}
In this step, voter $n$ wants to share  its vote $\mathbf{V}_n$, which is a one-hot vector in $\{0,1\}^{K+1} $, i.e., if voter $n$ votes to $C_k$, then $\mathbf{V}_n$ is equal to $\mathbf{e}_k$. Let us define $\mathbf{V'}_n$ as the complement of $\mathbf{V}_n$, equivalently, $\mathbf{V'}_n=[1,1,\dots,1]^T_{1\times(K+1)}-\mathbf{V}_n$ . Note that vector $\mathbf{V}_n$ has a single $1$ entity, and all the other entities are $0$. Thus, $\mathbf{V'}_n$ is a one-cold vector, i.e., its entities are $1$, except a single $0$.

In this step, voter $n$ shares both $\mathbf{V}_n$ and $\mathbf{V'}_n$ using verifiable secret sharing algorithm~\cite{chor1985verifiable}.
In order to do that voter $n$ constructs polynomials $\mathbf{F}^{(n)}(x)=\mathbf{V'}_n+\mathbf{R}^{(n)}_{1}x+\mathbf{R}^{(n)}_{2}x^2+\dots+\mathbf{R}^{(n)}_{t}x^t$ and $\mathbf{G}^{(n)}(x)=\mathbf{V}_n+\mathbf{Z}^{(n)}_{1}x+\mathbf{Z}^{(n)}_{2}x^2+\dots+\mathbf{Z}^{(n)}_{t}x^t$, then sends
$\mathbf{F}^{(n)}(\alpha_{n'})$ and $\mathbf{G}^{(n)}(\alpha_{n'})$ to voter $n'$, $\forall n,n' \in [N]$, 	
where $\mathbf{R}^{(n)}_{j}$ and $\mathbf{Z}^{(n)}_{j}$ are chosen uniformly and independently at random from the field $\mathbb{F}^{K+1}$, $\forall j \in [t]$. Also, distinct $\alpha_{1},\alpha_{2},\dots,\alpha_{N}$ are chosen uniformly and independently at random from the field $\mathbb{F}$, and they are known by all the voters.

Using $\mathsf{VSS}$  ensures the voters that if $N \geq 3t+1$, shared values by the voter $n$ are consistent, i.e., they are lying on a polynomial of degree $t$, otherwise, honest (not malicious) voters can identify malicious voters who have adversarial behavior and omit them from the remaining part of algorithm \cite{chor1985verifiable}. It must be mentioned that if voter $n$ is honest, the other voters cannot gain any information about the secret of $\mathbf{F}^{(n)}(x)$ and $\mathbf{G}^{(n)}(x)$, i.e., the values $\mathbf{F}^{(n)}(0)$ and $\mathbf{G}^{(n)}(0)$.

\subsection{Verification}
\label{subsection:PVSverification}
In this step, each voter needs to be assured that $\forall n \in [N]$, $\mathbf{V}_n$ is a one-hot vector. Actually, it satisfies the voters that 	voter $n$ follows the protocol and votes to exactly one of the candidates. In order to do that, we propose a 3-phase verification:\\ 1) Verification of summation: All voters verify whether $\mathbf{F}^{(n)}(0) + \mathbf{G}^{(n)}(0)$ is equal to $\mathbf{1}_{(K+1)\times 1}$ or not, $\forall n \in [N]$. 2) Verification of product: Voters verify whether $\mathbf{F}^{(n)}(0)*\mathbf{G}^{(n)}(0)=[\mathbf{F}^{(n)}(0)_1\mathbf{G}^{(n)}(0)_1,\mathbf{F}^{(n)}(0)_2\mathbf{G}^{(n)}(0)_2,\dots,$ $\mathbf{F}^{(n)}(0)_{K+1}\mathbf{G}^{(n)}(0)_{K+1}]^T$ is equal to $\mathbf{0}_{(K+1)\times 1}$ or not, $\forall n \in [N]$.\\
3) Verification of entities: voter $n$ broadcasts which polynomial between $\mathbf{F}^{(n)}(x),\mathbf{G}^{(n)}(x)$ is its vote. For simplicity, voter $n$ broadcasted  polynomial is denoted by $\mathbf{V}^{(n)}(x)$. Then, other voters must be assured that Sum$(\mathbf{V}^{(n)})=\displaystyle\sum_{i=1}^{K+1}V^{(n)}_i=1$.

If the first two conditions are satisfied, then we can conclude that $\{\mathbf{F}^{(n)}(0),\mathbf{G}^{(n)}(0)\}\in\{0,1\}^{K+1}$. Then, the last condition ensures the other voters that $\mathbf{V}^{(n)}(0)=\mathbf{V}^{(n)}=[V^{(n)}_1,V^{(n)}_2,\dots,V^{(n)}_{K+1}]$ is a one-hot vector. Thus, if all of the aforementioned conditions are satisfied, then we can conclude that the vote of voter $n$ is valid, i.e., voter $n$ votes to one of the candidates or abstain, $\forall n \in [N]$.

\subsubsection[bold]{\textbf{Verification of summation}} Define $\mathbf{S}^{(n)}(x) \defeq \mathbf{F}^{(n)}(x)+\mathbf{G}^{(n)}(x)$. In this phase, $\forall n,n' \in [N]$, voter $n'$ broadcasts $\mathbf{S}^{(n)}(\alpha_{n'})=\mathbf{F}^{(n)}(\alpha_{n'})+\mathbf{G}^{(n)}(\alpha_{n'})$. If all of the voters are honest, after this phase, each voter would have access to $\{\mathbf{S}^{(n)}(\alpha_{1}),\mathbf{S}^{(n)}(\alpha_{2}),\dots,\mathbf{S}^{(n)}(\alpha_{N})\}$. Noting $\deg(\mathbf{S}^{(n)}(x))=t$ and considering up to $t$ malicious voters, as long as $N\geq 3t+1$,  due to the Remark~\ref{remark:reedsolomon}, each voter can recover the correct set of $\{\mathbf{S}^{(n)}(\alpha_{1}),\mathbf{S}^{(n)}(\alpha_{2}),\dots,\mathbf{S}^{(n)}(\alpha_{N})\}$, $\forall n\in [N]$. As a consequence, each voter can calculate $\mathbf{S}^{(n)}(x)$, then, derive $\mathbf{S}^{(n)}(0) = \mathbf{F}^{(n)}(0)+\mathbf{G}^{(n)}(0)$, and finally verify whether $\mathbf{F}^{(n)}(0) + \mathbf{G}^{(n)}(0)$ is equal to $\mathbf{1}_{(K+1)\times 1}$ or not, $\forall n \in [N]$.
	
\subsubsection{\textbf{Verification of Product}} In this phase, each voter verifies whether $\mathbf{F}^{(n)}*(0)\mathbf{G}^{(n)}(0)$ is equal to $\mathbf{0}_{(K+1)\times 1}$ or not, $\forall n \in [N]$. In order to do that, according to Theorem \ref{theorem:Oiha}, each voter~$n$ finds polynomials $\mathbf{O}^{(n)}_1(x),\mathbf{O}^{(n)}_2(x),\dots,\mathbf{O}^{(n)}_t(x)$, such that $\deg( \mathbf{F}^{(n)}(x)*\mathbf{G}^{(n)}(x)-\displaystyle\sum_{i=1}^{t}x^i \mathbf{O}^{(n)}_i(x)) \leq t$. Let us define	
	\begin{align}
	\label{eq:PVSmult}
	\mathbf{C}^{(n)}(x) \defeq \mathbf{F}^{(n)}(x)*\mathbf{G}^{(n)}(x)-\displaystyle\sum_{i=1}^{t}x^i \mathbf{O}^{(n)}_i(x).
	\end{align}
	One can see that $\mathbf{C}^{(n)}(0)=\mathbf{F}^{(n)}(0)*\mathbf{G}^{(n)}(0)$. This is due to the fact that each $\mathbf{O}^{(n)}_i(x)$ is multiplied by $x^i$, where $i \geq 1$. Thus, the constant term of $\mathbf{F}^{(n)}(x)*\mathbf{G}^{(n)}(x)$ is not affected by $\mathbf{O}^{(n)}_i(x)$, $\forall i \in [t]$ and $\forall n\in [N]$.

	After constructing $\mathbf{O}^{(n)}_1(x),\mathbf{O}^{(n)}_2(x),\dots,\mathbf{O}^{(n)}_t(x)$, voter $n$ shares $\mathbf{O}^{(n)}_i(x)$ with all other voters by using $\mathsf{VSS}$ algorithm, i.e., it sends $\mathbf{O}^{(n)}_{i}(\alpha_{n'})$ to voter $n'$, $\forall n,n' \in [N]$, and  $\forall i \in [t]$. Also, voter $n$ shares $\mathbf{C}^{(n)}(x)$, i.e., it sends $\mathbf{C}^{(n)}(\alpha_{n'})$ to voter $n'$, $\forall n,n' \in [N]$. Until now, voter $n'$ has the values of $\mathbf{C}^{(n)}(x),\mathbf{F}^{(n)}(x),\mathbf{G}^{(n)}(x),\mathbf{O}^{(n)}_i(x)$ at point $\alpha_{n'}$, $\forall n,n' \in [N]$ and $\forall i \in [t]$. Hence, voter $n'$ can directly verify whether \eqref{eq:PVSmult} is held at $\alpha_{n'}$ or not. If \eqref{eq:PVSmult} is not held at $\alpha_{n'}$, then voter $n'$ broadcasts $\mathsf{Complaint}$ messages. Similar to  verification step in Section~\ref{section:motivating}, the other voters can compute the values $\mathbf{C}^{(n)}(\alpha_{n'}),\mathbf{F}^{(n)}(\alpha_{n'}),\mathbf{G}^{(n)}(\alpha_{n'}),\mathbf{O}^{(n)}_1(\alpha_{n'}),$ $\mathbf{O}^{(n)}_2(\alpha_{n'}),\dots,\mathbf{O}^{(n)}_t(\alpha_{n'})$ in collaboration with each other, to identify the malicious voter among voters $n$ and $n'$ and omit the malicious one from the remaining part of the algorithm.
	To be more precise, one can see \cite{fullproof}.

	Then, each voter $n'$ broadcasts the value of $\mathbf{C}^{(n)}(\alpha_{n'})$, and as a result, each voter has access to the value of $\mathbf{C}^{(n)}(x)$ at more than $3t$ points. Thus, due to the Remark~\ref{remark:reedsolomon}, each voter can compute $\mathbf{C}^{(n)}(x)$ and verify whether the value of $\mathbf{C}^{(n)}(0)=\mathbf{F}^{(n)}(0)*\mathbf{G}^{(n)}(0)$ is equal to 0 or not.

\subsubsection{\textbf{Verification of entities}} In this phase, voter $n$ broadcasts its vote, i.e., it broadcasts  the secret of which polynomial between $\mathbf{F}^{(n)}(x)$ and $\mathbf{G}^{(n)}(x)$ is its vote.
The aim is to verify that sum$(\mathbf{V}^{(n)}(0))=$ sum$(\mathbf{V}^{(n)})=1$. 
	
	In order to do that, $\forall n,n' \in [N]$, each voter $n'$ broadcasts the value sum$(\mathbf{V}^{(n)}(\alpha_{n'}))$. If all of the voters were honest, after this phase each voter has access to $\{\text{Sum}(\mathbf{V}^{(n)}(\alpha_{1})),\text{Sum}(\mathbf{V}^{(n)}(\alpha_{2}))\dots,$ $\text{Sum}(\mathbf{V}^{(n)}(\alpha_{N}))\}$, which are located on the $t-$degree polynomial $\text{Sum}(\mathbf{V}^{(n)}(x))$. However, in reality, some of the voters are malicious and do adversarial behavior. As it is mentioned in Remark~\ref{remark:reedsolomon}, voters can correct up to $\frac{N-t}{t}$ errors, or equivalently, if $N\geq 3t+1$, each voter can calculate Sum$(\mathbf{V}^{(n)}(0))$, then derive Sum$(\mathbf{V}^{(n)})$.
	
The first two phases confirm that $\mathbf{V}_n \in \{0,1\}^{K+1}$. The third condition ensures the other voters that $\mathbf{V}^{(n)}(0)=\mathbf{V}^{(n)}=[V^{(n)}_1,V^{(n)}_2,\dots,V^{(n)}_{K+1}]$ is a one-hot vector.

\subsection{Counting}
\label{subsection:PVScounting}

Assume that $\mathcal{I}$ is the set of all malicious voters that are identified by the other voters. So far, each voter $n'$ has $\mathbf{V}^{(n)}(\alpha_{n'})$ . Also, all of the voters are assured that $\mathbf{V}^{(n)}$ is a one-hot vector, i.e, exactly one of the entities of $\mathbf{V}^{(n)}$ is equal to 1 and the other entities are equal to 0. In this step, voter $n'$ computes $\mathbf{V}(\alpha_{n'})= \displaystyle\sum_{n \in [N]\backslash \mathcal{I}} \mathbf{V}^{(n)}(\alpha_{n'})$ and broadcasts the result. Ideally, after this step, each voter has access to $\{\mathbf{V}(\alpha_{1}),\mathbf{V}(\alpha_{2}),\dots,\mathbf{V}(\alpha_{N})\}$. But, considering malicious voters that have adversarial behavior, by using Reed-Solomon decoding procedure, and the fact that $\deg(\mathbf{V}(x))=t$, voters can correct up to $\frac{N-t}{2}$ errors. Since the number of malicious voters who are omitted or not is at most $t$, we need to have $\frac{N-t}{2}\geq t$, or equivalently, $N\geq 3t+1$.
If $N\geq 3t+1$, each voter can recover the correct set of $\{\mathbf{V}(\alpha_{a_1}),\mathbf{V}(\alpha_{a_2}),\dots,\mathbf{V}(\alpha_{a_{|[N]\backslash\mathcal{I}|}})\}$, $\forall \alpha_i \in [N] \backslash \mathcal{I}$, 
and calculate $\mathbf{V}(x)$, and finally derive $\mathbf{V}(0) = \displaystyle\sum_{n \in [N]\backslash \mathcal{I}}\mathbf{V}^{(n)}(0)=\displaystyle\sum_{n \in [N]\backslash \mathcal{I}}\mathbf{V}_n$, which is equal to our final result $\mathbf{R}=[R_1,R_2,\dots,R_{K+1}]^T$ casted in our voting, without counting the votes of identified malicious voters in set $\mathcal{I}$ .

As described above as long as $N\geq 3t+1$, the correctness and robustness conditions are satisfied. The privacy is assured using verifiable secret sharing, the proof is similar to Appednix \ref{sec:privacy}.

\section{Conclusion} \label{sec::conclusion}

In this paper, we propose an information-theoretic secure and private voting system. We use multi-party computation and verifiable secret sharing to detect, correct, or drop malicious voters. It is shown that if the total number of voters is greater than three times of malicious voters, then the system can handle adversarial behavior.
An interesting future research direction is to expand the voting system to satisfy other conditions besides correctness, privacy, and robustness.

\bibliographystyle{ieeetr}
\bibliography{journal_abbr,polynomialMPC} 

\appendices

\section{Privacy Overview of Motivating Example}
\label{sec:privacy}
Here, we prove the privacy of the proposed scheme in Section \ref{section:motivating}, for the case where there are exactly $t$ malicious voters. On the other hand, in the case of less than $t$ malicious voters, the set of all massages that the malicious voters received is a subset of the first case. Thus, we just prove the first case.  For simplicity, assume that the set of malicious voters are voter 1, voter 2, ..., and voter $t$. In the following, we prove the privacy of  motivating example step by step.

\subsection{Sharing}
\label{subsec:privacysharing}

In this step, the set of malicious voters has access to the values of $\mathbf{F}^{(n)}(x)$ and $\mathbf{G}^{(n)}(x)$ at $t$ different points $\alpha_1 , \dots, \alpha_t$, for all $n \in [N]$. Accordingly, they have access to the sets $\{\mathbf{F}^{(n)}(\alpha_{1}),\mathbf{F}^{(n)}(\alpha_{2}),\dots,\mathbf{F}^{(n)}(\alpha_{t})\}$ and $\{\mathbf{G}^{(n)}(\alpha_{1}),\mathbf{G}^{(n)}(\alpha_{2}),\dots,$ $\mathbf{G}^{(n)}(\alpha_{t})\}$. Through information-theoretic privacy of Shamir secret sharing \cite{shamir1979share}, since the malicious voters have access to $t$ points of polynomial of degree $t$, they get no information about that. If they had another point on these polynomials, they could uniquely determine both of them. 

For an honest voter $n$, the constant term of $\mathbf{F}^{(n)}(x)$ is 1 or 0. Thus, in the malicious voters' point of view, there exist exactly two possible candidates for $\mathbf{F}^{(n)}(x)$,  which are shown by $\mathbf{F}^{(n,0)}(x)$  and $\mathbf{F}^{(n,1)}(x)$ such that $\mathbf{F}^{(n,0)}(0)=0$  and $\mathbf{F}^{(n,1)}(0)=1$. Similarly, $\mathbf{G}^{(n,0)}(x)$  and $\mathbf{G}^{(n,1)}(x)$ can be defined.

\subsection{Verification}
\label{subsec:privacyverification}
In this step, we follows a 2-phase scheme:

\textbf{Verification of summation:}
In this phase, each voter $n'$ broadcasts the value $\mathbf{S}^{(n)}(\alpha_{n'})=\mathbf{F}^{(n)}(\alpha_{n'})+\mathbf{G}^{(n)}(\alpha_{n'}),$ $\forall n,n'\in [N]$. For each honest voter $n$, the value of $\mathbf{F}^{(n)}(0)+\mathbf{G}^{(n)}(0)$ is equal to $1$. Also, malicious voters have the value of $\mathbf{S}^{(n)}(x)$ at $t$ different points $\{\alpha_{1},\alpha_{2},\dots,\alpha_{t}\}$ in advance. Accordingly, they could already compute $\mathbf{S}^{(n)}(x)$. Thus, they cannot gain any additional information in this phase.

\textbf{Verification of Product:}
In this phase, malicious voters receive the values of $\mathbf{C}^{(n)}(x),\mathbf{O}^{(n)}_1(x),$ $\mathbf{O}^{(n)}_2(x),\dots,\mathbf{O}^{(n)}_t(x)$ at $t$ different points $\{\alpha_{1},\alpha_{2},\dots,\alpha_{t}\}$.
One can see that there exist polynomials $\mathbf{O}^{(n,0)}_1(x),\mathbf{O}^{(n,0)}_2(x),\dots,\mathbf{O}^{(n,0)}_t(x)$ of degree $t$ such that $\mathbf{O}^{(n,0)}_i(\alpha_{j})=\mathbf{O}^{(n)}_i(\alpha_{j}), \forall i,j \in [t]$ and $\deg( \mathbf{F}^{(n,0)}(x)*\mathbf{G}^{(n,1)}(x)-\displaystyle\sum_{i=1}^{t}x^i \mathbf{O}^{(n,0)}_i(x)) \leq t$. Also there exist degree-$t$ polynomials $\mathbf{O}^{(n,1)}_1(x),\mathbf{O}^{(n,1)}_2(x),\dots,\mathbf{O}^{(n,1)}_t(x)$ such that $\mathbf{O}^{(n,1)}_i(\alpha_{j})=\mathbf{O}^{(n)}_i(\alpha_{j}), \forall i,j \in [t]$ and $\deg( \mathbf{F}^{(n,1)}(x)*\mathbf{G}^{(n,0)}(x)-\displaystyle\sum_{i=1}^{t}x^i \mathbf{O}^{(n,1)}_i(x)) \leq t$. Thus, malicious voters cannot distinguish $\{\mathbf{F}^{(n,1)}(x),\mathbf{G}^{(n,0)}(x)\}$ from $\{\mathbf{F}^{(n,0)}(x),\mathbf{G}^{(n,1)}(x)\}$ to be able to derive the main polynomials $\{\mathbf{F}^{(n)}(x),\mathbf{G}^{(n)}(x)\}$.

Then, each voter $n'$ broadcasts $\mathbf{C}^{(n)}(n')$. For each honest voter $n$, $\mathbf{C}^{(n)}(0)= \mathbf{F}^{(n)}(0).\mathbf{G}^{(n)}(0)$ which is equal to~$0$. Also, malicious voters already have the value of $\mathbf{C}^{(n)}(x)$ at $t$ different points $\{\alpha_{1},\alpha_{2},\dots,\alpha_{t}\}$. Accordingly, they can compute $\mathbf{C}^{(n)}(x)$. Thus, broadcasting $\mathbf{C}^{(n)}(x)$ at different points does not add any additional information to the malicious voters.

\subsection{Counting:}
\label{subsec:privacycounting}
In this step, each voter $n'$ broadcasts $\mathbf{V}(\alpha_{n'})= \displaystyle\sum_{n \in [N]} \mathbf{V}^{(n)}(\alpha_{n'})$. Then, each voter is able to compute polynomial $\mathbf{V}(x)$ and gain $\mathbf{V}(0)$ which is equal to the total number of $\mathsf{YES}$ votes. Assume that the total number of $\mathsf{YES}$ votes is $\mathsf{Y}$. 
One can see that $\mathbf{Q}_1(x) \defeq \mathbf{G}^{(i,1)}(x)+\mathbf{F}^{(j,0)}(x)$ is equal to $\mathbf{Q}_2(x)\defeq \mathbf{G}^{(i,0)}(x)+\mathbf{F}^{(j,1)}(x)$. This is due to the fact that, $\mathbf{Q}_1(0)=\mathbf{Q}_2(0)=1$ and equality of  $\mathbf{G}^{(i,1)}(x)+\mathbf{F}^{(j,0)}(x)$ and  $\mathbf{G}^{(i,0)}(x)+\mathbf{F}^{(j,1)}(x)$ is hold at $t$ different point $\alpha_{1},\alpha_{2},\dots,\alpha_{t}$. It means that, there is no difference in computing $\mathbf{V}(x)$, if voter $i$ votes $\mathsf{YES}$ and voter $j$ votes $\mathsf{No}$  or vice versa.
Thus, malicious voters cannot determine that which of the voters vote $\mathsf{YES}$ and which ones vote $\mathsf{No}$.

\end{document}